\documentclass{nature}

\usepackage{float}
\usepackage{dcolumn}
\usepackage{bm}
\usepackage[colorlinks=true]{hyperref}
\usepackage{color}
\usepackage{array}
\usepackage{multirow}
\usepackage{amssymb}
\usepackage[normalem]{ulem}
\usepackage{amsmath}
\usepackage{xcolor}
\usepackage{soul}
\usepackage{graphicx}
\makeatletter
\let\saved@includegraphics\includegraphics
\AtBeginDocument{\let\includegraphics\saved@includegraphics}
\renewenvironment*{figure}{\@float{figure}}{\end@float}
\makeatother

\definecolor{RED}{rgb}{1,0,0}

\def\cob#1{compact object binaries#1
  (COB#1)\gdef\cob{COB}}
\def\imbh#1{intermediate-mass black hole#1
  (IMBH#1)\gdef\imbh{IMBH}}
\def\imbhb#1{intermediate-mass black hole binary#1
  (IMBHB#1)\gdef\imbhb{IMBHB}}  
\def\bh#1{black hole#1
  (BH#1)\gdef\bh{BH}}
\def\bbh#1{binary black hole#1
  (BBH#1)\gdef\bbh{BBH}}
\def\gw#1{gravitational wave#1
  (GW#1)\gdef\gw{GW}}
\def\nr#1{numerical relativity#1
<argument> \mnras  (NR#1)\gdef\nr{NR}}
\def\snr#1{signal-to-noise-ratio#1
  (SNR#1)\gdef\snr{SNR}}
\def\pn#1{post-Newtonian#1
  (PN#1)\gdef\pn{PN}}
  \def\O1#1{observing run#1
  (O1#1)\gdef\pn{O1}}

\title{Detectability of Intermediate-Mass Black Holes in \\
Multiband Gravitational Wave Astronomy}

\author{Karan Jani $^{1}$\footnote{Corresponding author}, Deirdre Shoemaker $^{1}$, Curt Cutler $^{2,3}$}

\begin{document}

\maketitle

\begin{affiliations}
 \item Center for Relativistic Astrophysics, School of Physics, Georgia Institute of Technology, 837 State St., Atlanta, GA, USA - 30363 
 \item Theoretical Astrophysics, California Institute of Technology, Pasadena, CA 91125, USA
 \item Jet Propulsion Laboratory, 4800 Oak Grove Drive, Pasadena, CA 91109, USA
\end{affiliations}

\begin{abstract}
The direct measurement of gravitational waves is a powerful tool for surveying the population of black holes across the universe. The first gravitational wave catalog from LIGO\cite{O2Catalog} has detected black holes as heavy as $\sim50~M_\odot$, colliding when our Universe was half its current age. However, there is yet no unambiguous evidence of black holes in the intermediate-mass range of $10^{2-5}~M_\odot$. Recent electromagnetic observations have hinted at the existence of \imbh{s} in the local universe\cite{2009Natur.460...73F, 2014Natur.513...74P, 2017Natur.542..203K}; however their masses are poorly constrained. The likely formation mechanisms of \imbh{s} are also not understood. Here we make the case that multiband gravitational wave astronomy\cite{2016PhRvL.116w1102S, WP-Multiband} --specifically, joint observations by space- and ground-based gravitational wave detectors--will be able to survey a broad population of \imbh{s} at cosmological distances. By utilizing general relativistic simulations of merging black holes\cite{JANI_2016} and state-of-the-art gravitational waveform models\cite{PhysRevLett.120.161102}, we classify three distinct population of binaries with \imbh{s} in the multiband era and discuss what can be observed about each. Our studies show that multiband observations involving the upgraded LIGO detector\cite{LIGO-Instrument2018} and the proposed space-mission LISA\cite{2017arXiv170200786A} would detect the inspiral, merger and ringdown of \imbh{} binaries out to redshift $z \sim 2$. Assuming that next-generation detectors, Einstein Telescope\cite{Punturo:2010zz} and Cosmic Explorer\cite{2019arXiv190704833R}, are operational during LISA's mission lifetime, we should have multiband detections of \imbh{} binaries out to redshift $z \sim 5$. To facilitate studies on multiband \imbh{} sources, here we investigate the multiband detectability of \imbh{} binaries. We provide analytic relations for the maximum redshift of multiband detectability, as a function of black hole mass, for various detector combinations.
Our study paves the way for future work on what can be learned from \imbh{} observations in the era of multiband gravitational wave astronomy.


\end{abstract}

\begin{figure}
\centering
 \includegraphics[scale=0.42,trim = {60 0 0 0}]{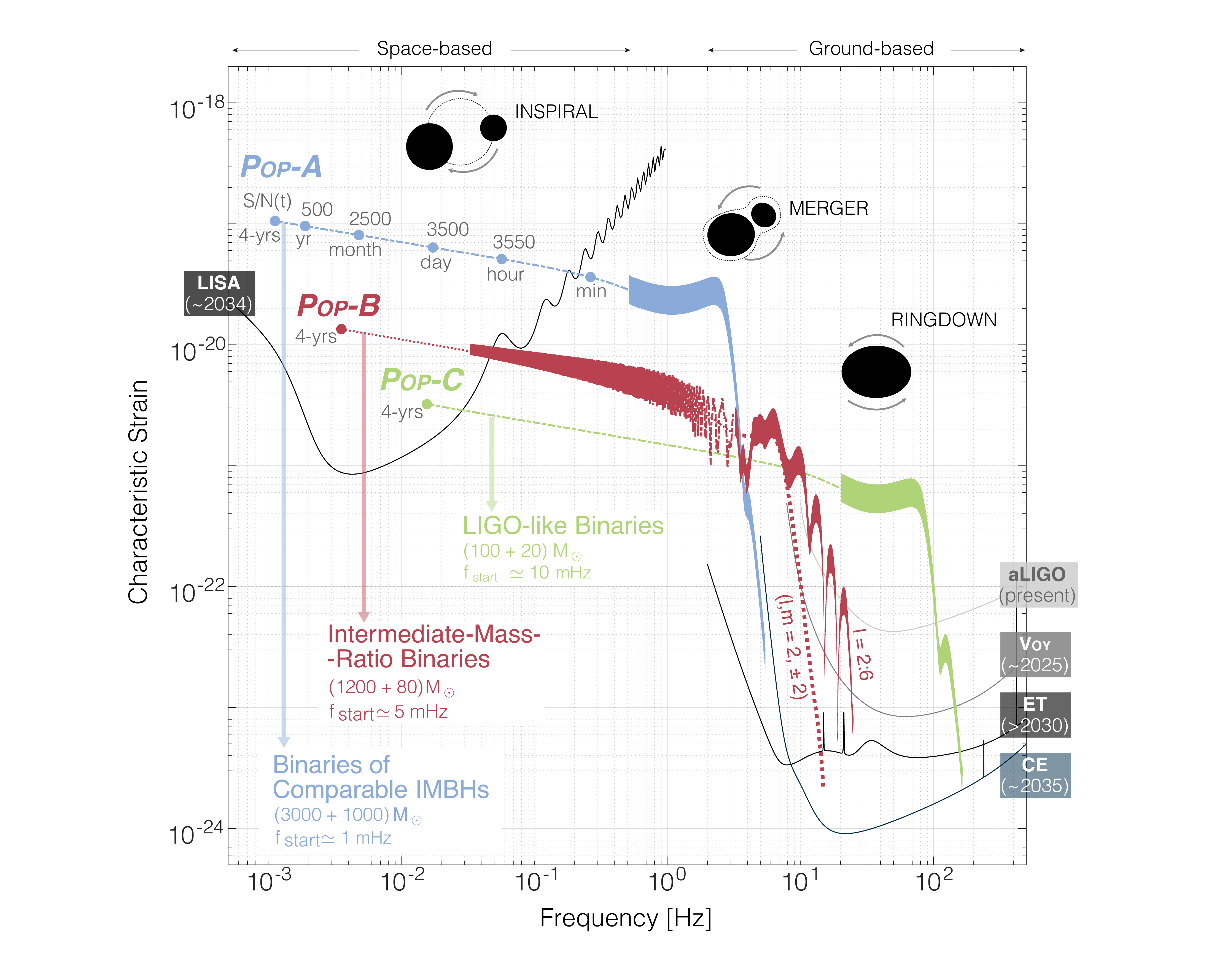} 
\caption{{\bf Examples of binaries with \imbh{s} in multiband gravitational wave spectrum}. (Continued on the following page).
}
\label{Fig1} 
\end{figure}

Two black holes in a gravitationally bound system typically orbit around their joint center of mass for millions of years, emitting gravitational waves of monotonically increasing frequency, before ultimately colliding. These black hole collisions are the most energetic phenomena in the known universe. The data collected from LIGO-like detectors provide the most stringent bounds on the population of such merging black hole binaries as a function of their total mass. Ongoing observations show that stellar-mass black hole binaries ($15\sim100~M_\odot$) merge at least once per month per cubic Gpc\cite{O2Catalog}. Mergers of more massive binaries ($100-800~M_\odot$) are more rare, with the current upper-limit on their merger rate being one per few years per cubic Gpc\cite{O2IMBH}. The latter is the strongest observational constraint on the merger rate of lower-range \imbh{s}: $10^{2-3}~M_\odot$. Mergers of medium- ($10^{3-4}~M_\odot$) and upper-range \imbh{s} ($10^{4-5}~M_\odot$) are outside the sensitivity band of current generation detectors. 
We divide \imbh{s}, by mass range, into three distinct binary populations (as shown in figure-\ref{Fig1} and -\ref{Fig4}): {\it Pop-A} (binaries with \imbh{s} of comparable masses), {\it Pop-B} (intermediate-mass-ratio binaries), {\it Pop-C} (binaries with LIGO-like remnants) and determine out to what distance these populations will be jointly detectable by gravitational wave detectors in both the milli-Hz (space-based missions) and hecto-Hz (ground-based detectors) bands.

\begin{figure}
\par
{Figure 1: We define the multiband detection of a binary as the measurement of its inspiral in LISA throughout the 4-year mission lifetime ($0.001-0.1$ Hz) and at least the merger or ringdown in the ground-based detectors (\texttt{ET, CE, VOY}: $1-250$ Hz). The y-axis is the characteristic strain\cite{2015CQGra..32a5014M}, which is defined as $2f|h(f)|$ for the binary black holes and $\sqrt{f~S_n}$ for the instrumental noise. The binary evolution is shown for examples of Pop-A,B,C at redshift $z=0.5$. Pop-B is placed at an inclination of $\pi/4$, while Pop-A,C are face-on. The late-inspiral, merger and ringdown of Pop-A,B,C were computed with numerical relativity simulations\cite{JANI_2016} (thick lines), while the early inspiral (dotted-dashed lines) were computed using \texttt{IMRPhenomHM}\cite{PhysRevLett.120.161102}. Both models includes higher-order harmonics of gravitational radiation $(\ell \geq 2)$. The dominant mode $(\ell, m) = (2,\pm2)$ is shown in dotted line for Pop-B. Here $f_\text{start}$ is set for a 4-year LISA mission. The evolution of signal-to-noise-ratio $(S/N)$ over time $t$ spent in the LISA band is highlighted for Pop-A. 
}
\end{figure}

\begin{figure}
\centering
 \includegraphics[scale=0.38,trim = {20 130  0 50 }]{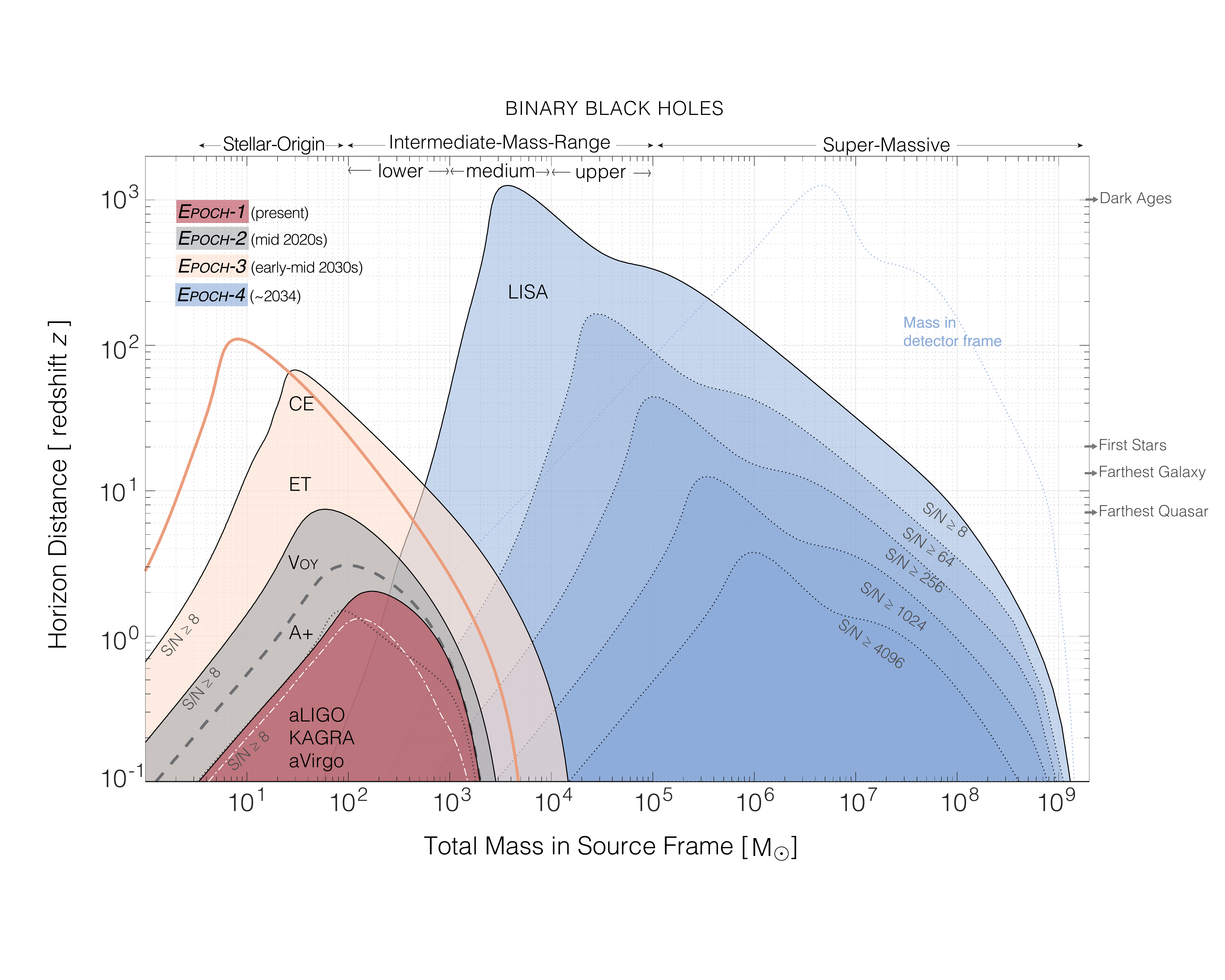} 
\caption{{\bf Cosmological reach to black hole binaries within the next 20 years of gravitational wave astronomy}. 
The x-axis displays the total-mass measured in the source frame.  
The total-mass measured in the detector frame is heavier by $(1+z)$ (as shown by dotted line for LISA). In this paper we consider binaries with total-mass in the \imbh{} range ($10^{2-5}~M_\odot$). We divide \imbh{} binaries into three classes - (i) lower-range $10^{2-3}~M_\odot$, (ii) medium-range $10^{3-4}~M_\odot$, and (iii) upper-range $10^{4-5}~M_\odot$ masses. The horizon distance (measured as redshift) on the y-axis has been computed at the minimum threshold $(S/N) = 8$ for binaries with two equal-sized, non-spinning black holes. For LISA, we  show how the horizon radius depends on $S/N$. See methodology for further details about instrumental noise.
}        
\label{Fig2} 
\end{figure}

Figure-\ref{Fig2} provides an overview of the binary black hole mass-range - from stellar to intermediate to supermassive - that will be surveyed during the next 20 years of gravitational wave astronomy. For discussion in this text, we classify four epochs of gravitational wave experiments (see Supplemental Material for additional details). Epoch-1 (present) is the current generation of Advanced LIGO-like detectors and epoch-2 (early to mid-2020s) will be the upgrades to the existing LIGO facilities called A+ and Voyager (\texttt{Voy}). Epoch-3 (early to mid-2030s) may consists of two 3rd generation detectors - Einstein Telescope (\texttt{ET})\cite{Punturo:2010zz} and Cosmic Explorer (\texttt{CE})\cite{Evans:2016mbw}. Epoch-4 ($\sim2034$) includes space-based gravitational wave detectors such as LISA\cite{2017arXiv170200786A} and perhaps other concept designs\cite{2018arXiv181104743L}. Each successive epoch of detectors improves sensitivity for low-frequency gravitational waves. This in turn allows us to detect signals from heavier \imbh{s}, as the frequency at the merger scales inversely with total-mass. The luminosity distance to which we can find \imbh{s} in each epoch primarily depends on their masses. Unless stated otherwise, all the masses quoted in this text are measured in their source frame, and distances refer to luminosity distance. Gravitational wave measurements of binaries provide stringent constraints on their redshifted chirp-mass:\cite{1994PhRvD..49.2658C} $\mathcal{M}_c = \eta^{3/5}~M_\mathrm{src}~(1+z)$, where $M_\mathrm{src}$ is the total mass of the binary located at redshift $z$, and the symmetric mass-ratio $\eta$ is defined by
$\eta = m_1~m_2/(m_1+m_2)^2$. An observed chirp-mass of $\mathcal{M}_c \sim ~ 10^{2-4}~{M}_\odot$ may yield astronomy's first unambiguous detection of an \imbh{}.

For binaries with \imbh{s} in the lower and medium-range ($10^{2-4}~{M}_\odot$), LISA-like space-based detectors will measure only their early inspiral, from $1-100~\text{milli-Hz}$. This inspiral could last months to years depending on the mass of the black holes. However, the loudest gravitational wave signature - the merger of \imbh{s} - would occur in the frequency range accessible to LIGO-like ground-based detectors, $\sim 1-250$ Hz. A measurement of the binary's evolution in {\it both} the ground- and space-based detectors is what we define as a {\it multiband observation}. Our minimal criteria for a multiband detection of a single binary source are the observation of its inspiral by LISA throughout the mission's 4-year lifetime ($0.001-0.1$Hz), plus the observation of its merger or ringdown by at least one of the ground-based detectors (as shown for examples in figure-\ref{Fig1}). Additionally, our criteria for "multiband detection" includes that the source should be measured with $S/N\geq8$ in both ground- and space-based detectors . 

The principal motivation of our study is to chart out the parameter space of multiband observations with \imbh{s}. These are particularly exciting sources, as their multiband observations should permit stronger tests of general relativity (with separate measurement of the pre- and post-merger black holes)\cite{2016PhRvL.117e1102V}, tighter constraints on the formation channels of heavy black holes\cite{2019BAAS...51c.175B} (e.g., through better constraints on their spins) and their environment\cite{2019ApJ...878...75R} (by seeing the evolution of their orbital eccentricity). In this text, we investigate three multiband network combinations of LISA with upcoming ground-based detectors: \texttt{LISA + Voy}, \texttt{LISA + CE} and \texttt{LISA + ET}. The latter is the most favourable scenario as it offers the least gap between the two frequency bands.

For the simplest case of binaries with two equal-sized, non-spinning black holes, we derive analytic relations for multiband detection radius (measured as cosmological redshift $z_\text{multi}$). Depending on the total-mass in the source frame $M_\text{src} = M_\text{det}/(1+z_\text{multi})$ and detection threshold $S/N$, the multiband detection radius can be expressed as:
\begin{eqnarray}
z_{\text{multi}} \approx  \begin{cases}
               \vspace{0.3cm}
               \left[~a_1 \left(\dfrac{M_\text{det}}{100~{M}_\odot}\right)^{a_2}~ \right] \left(\dfrac{S/N}{8} \right)^{-1} \hspace{2.5cm} \text{for } M_{\text{src}} \text{ in stellar-mass range} \\
               \vspace{0.3cm}
                \left[~b_1 \left(\dfrac{M_\text{det}}{1000~{M}_\odot}\right)^{b_2} + b_3~\right] \left(\dfrac{S/N}{8} \right)^{-6/7}  \hspace{1.2cm} \text{for } M_{\text{src}}\text{ in lower-range IMBH} \\
              \left[ c_1~\exp{ \left( c_2 \dfrac{M_\text{det}}{M_0} \right)} \right] \left(\dfrac{S/N}{8} \right)^{-6/7}  \hspace{2.1cm} \text{for } M_{\text{src}}\text{ in medium-range IMBH} 
            \end{cases}
\label{eq-multiband}
\end{eqnarray}

\begin{figure}
\centering
 \includegraphics[scale=0.4,trim = {100 50 100 100}]{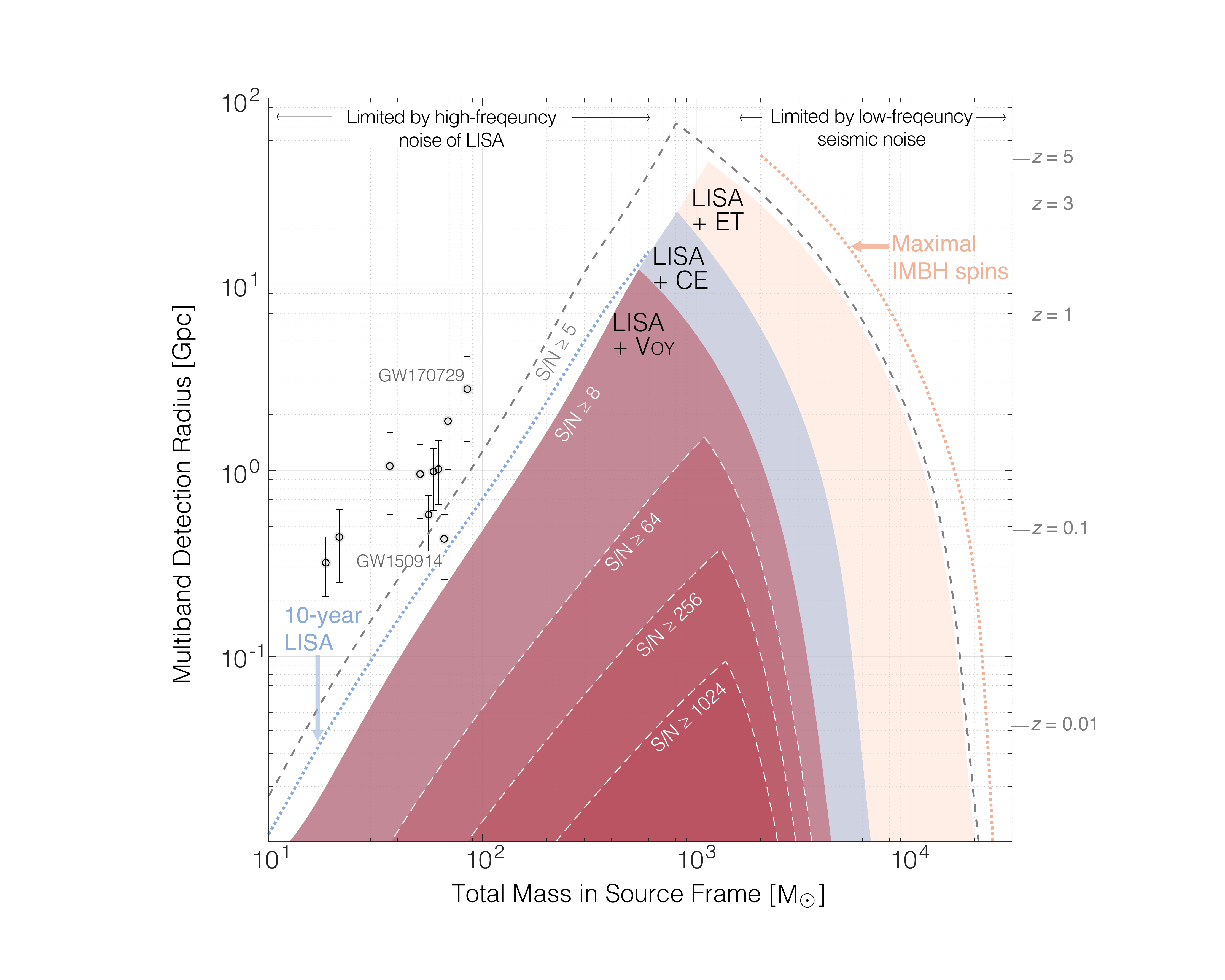}
\caption{ {\bf Multiband detection radius for black hole binaries}. The contours refer to binaries with two equal-sized, non-spinning black holes at detection threshold $S/N\geq 8$, for various multiband networks. The dashed white lines are for higher $S/N$, while the dashed black line is for a weaker detection limit of $S/N\geq5$. For \texttt{LISA + Voy}, the peak detection radius is 12 Gpc for $(260+260)~M_\odot$. While for \texttt{LISA + ET}, the peak is at 46 Gpc for $(570+570)~M_\odot$, and at 24 Gpc for $(405+405)~M_\odot$ in \texttt{LISA + CE}. The blue dotted line shows an increase in detection radius of $35\%$ for stellar black hole binaries with a 10-yr LISA mission, and the orange dotted line shows an increase of $40\%$ for upper-range \imbh{s}, with maximal, aligned spins. The black dots, shown with error bars in distance, refer to the black hole binaries reported in the Gravitational-Wave Transient Catalog-1\cite{O2Catalog}.  }
 
\label{Fig3} 
\end{figure}

In figure-\ref{Fig3}, we provide a diagrammatic version of eq. (\ref{eq-multiband}) at various $S/N$ levels, for three different multiband detector networks (see eq. (\ref{constants}) and Table-\ref{Table-p} for the constants). We find that \imbh{} binaries in the lower- to medium-range ($10^{2-4}~M_\odot$) are seen the farthest by all the multiband networks. For coalescences of $\sim10^3~M_\odot$ \imbh{} binaries within $1~\text{Gpc}^3$, all three multiband networks would detect them with $S/N\sim100$. Such a high $S/N$  in both bands should provide a unique opportunity to test the spin-orbit  coupling predicted by general relativity (by independently measuring spin evolution in the post-Newtonian regime with LISA and in the strong-gravity limit with ground-based detectors), stronger constraints on the Hubble constant \cite{2018MNRAS.475.3485D} (through tighter constraints on sky-location, inclination and distance for both ground- and space-based detectors) and by improving cross calibration of the space-based and ground-based instruments to better than $1\%$. 

For \imbh{s} with maximally aligned spins, we find the maximum detectable total-mass for multiband observations increases by a factor of $\sim1.5$, reaching up to $2.2\times10^4~M_\odot$ for the network \texttt{LISA + ET}. Further, the multiband detection radius can change by $\sim 75\%$ by effective spins of binaries with medium to upper-range IMBHs, but has virtually no impact for stellar-mass binaries and lower-range IMBHs (see figure-\ref{Fig6} and discussion in Supplemental Material). Further, we notice that the stellar-mass black hole binaries that LIGO has detected so far represent the weakest sources we are considering here, for all multiband networks. Except for GW150914 and GW170814, all other LIGO/Virgo events would be either sub-threshold or undetectable sources for LISA, assuming a 4-year LISA mission lifetime. Even for a 10-year lifetime, which would allow a lower $f_\text{start}$, the detection radius in LISA for stellar-mass black hole binaries increases by only $\sim30\%$.

We consider a wide range of binary mass distributions, to better delineate the parameter space accessible for multiband observations in the \texttt{LISA + ET} network in Figure-\ref{Fig4} (for other networks, see Figure-\ref{Fig5}  in Supplemental Material). We divide this parameter space into three distinct \imbh{} populations, labeled as Pop-A,B,C, which would potentially correspond to different formation astrophysical scenarios. Examples of these populations and their observation by multiband networks are shown in figure-\ref{Fig1}.

\paragraph{ Pop-A (Binaries whose \imbh{s} have comparable mass). } The inspirals from such binaries would be recorded in LISA starting at $f\gtrsim 1$ milli-Hz, accumulating strong $S/N$ during the 4-year mission lifetime. Similarly, a high $S/N$ from their mergers would be measured in the ground-based detectors. This makes Pop-A very promising sources for multiband observations. The coalescence of two medium-range IMBHs, $(3000+1000)~M_\odot$ (green curve in figure-\ref{Fig1}), would have a multiband detection radius of $\sim 8.5$~Gpc for the \texttt{LISA + ET} network. Lower-range IMBH binaries ($800 + 400)M_\odot$ will permit multiband detections up to $40$~Gpc  ($3.3$~Gpc) away for \texttt{LISA + ET} (\texttt{LISA + Voy}). Such cosmological reach should be particularly useful for probing the origins of early-forming \imbh{s} (such as remnants of Pop-III stars\cite{2001ApJ...551L..27M, 2019BAAS...51c.175B}). The multiband network of \texttt{LISA + ET} could detect mergers of ($10^4+2000)~M_\odot$ out to $400$ Mpc.
Further, we find that if such heavy black holes are spinning, this can increase the multiband detection radius by $\sim40\%$ (see figure-\ref{Fig6}).

\begin{figure}
\centering
 \includegraphics[scale=0.42,trim = {100 20 0 0}]{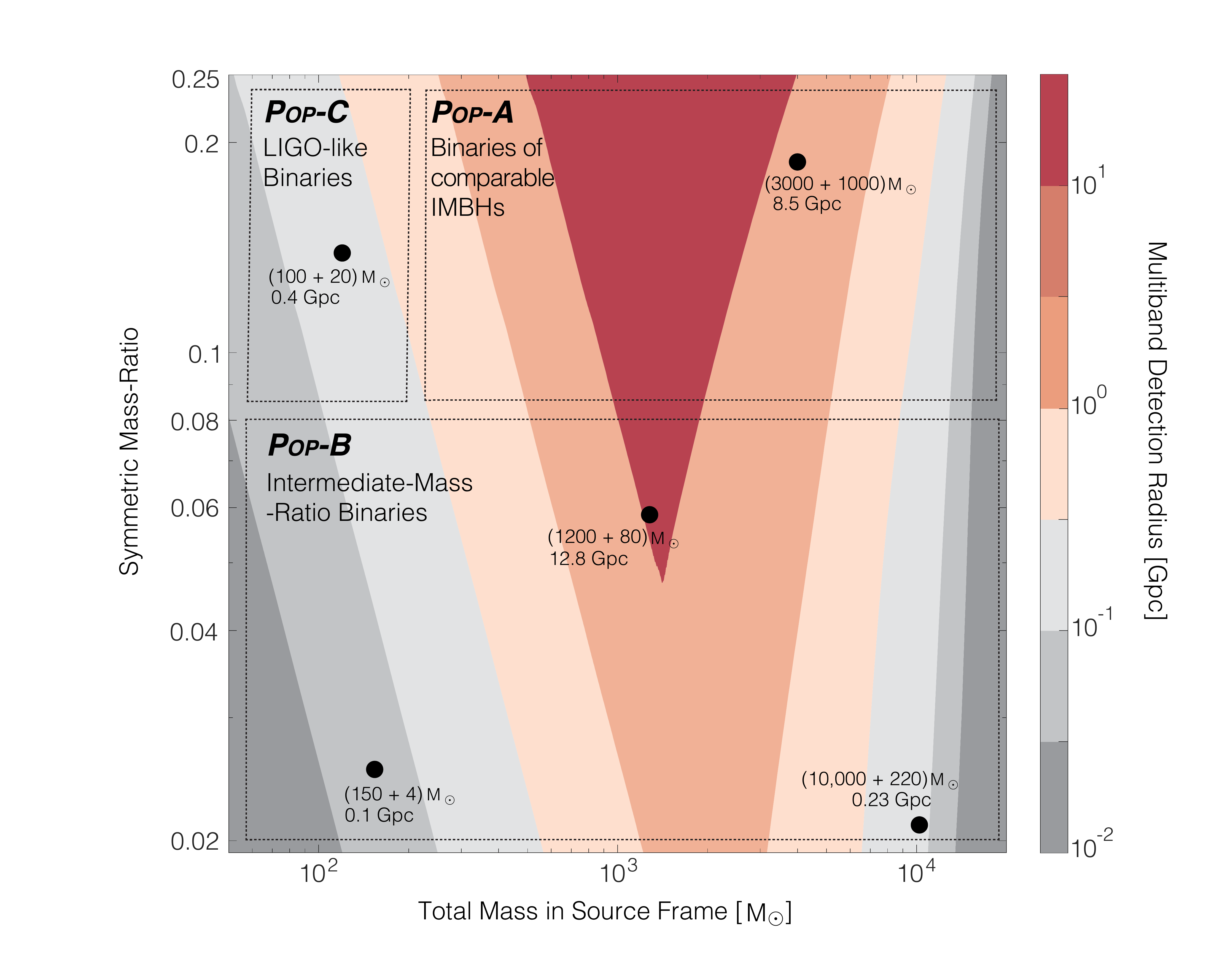} 
\caption{{\bf Parameter space of multiband black hole binaries in the \texttt{LISA + ET} network}. The y-axis plots the symmetric mass-ratio of the binary $(\eta \simeq 0.08$ for $m_1/m_2 = 1/10 $) and the contours are the multiband detection radius. The boundaries refer to the three distinct \imbh{} populations highlighted in figure-\ref{Fig1} and discussed in the text. Notice that independent of mass-ratio, the peak sensitivity remains for total-mass of $1.5$ to $2 \times 10^3~M_\odot$  }
\label{Fig4} 
\end{figure}
 
\paragraph{ Pop-B (Intermediate-mass-ratio binaries).} For such sources, we identify the primary object as an \imbh{} and the secondary object as a black hole such that the mass-ratio is $ < 1/10 $. The secondary could be either a stellar-mass black hole or lower-range \imbh{}. Pop-B inspirals would be observed by LISA from $f\gtrsim 5$ milli-Hz to $f\sim 0.1$ Hz during its 4-year mission lifetime. The energy radiated by the merger itself scales inversely with mass-ratio\cite{JANI_2016}, so unlike Pop-A, these sources would be faint or sub-optimal in the ground-based detectors. On the other hand, the small mass ratio entails significant gravitational wave emission in the higher-frequency, sub-dominant ($l > 2$) modes during the late-inspiral, merger and ringdown stages \cite{O1imbhHM}. Having acquired precise information from LISA regarding the source parameters (frequency and time of merger, orbital inclination angle) should facilitate measurement of these relatively lower-amplitude modes in ground-based detectors, up to $\ell\leq6$ (see red curve in figure-\ref{Fig1}). E.g., for the coalescence of a $80~M_\odot$ black hole with a $1200~M_\odot$ medium-range IMBH, we find that the sub-dominant modes would contribute about 75\% of the recorded $S/N$ in ground-based detectors. The \texttt{LISA + ET} network can measure $S/N\geq 100$ ($\geq 8$) for such a source within 0.4 Gpc (12.8 Gpc), allowing high resolution black hole spectroscopy and tests of the no-hair theorem in the multiband era\cite{2018arXiv180700075T}. 
In \texttt{LISA + Voy} network, a $S/N \sim 5$ would be possible for this source within 1 Gpc, thanks to the contribution from sub-dominant modes. 

For Pop-B sources with lower-range IMBH{s}, such as ($600+20)~M_\odot$, both \texttt{LISA + ET}  and \texttt{LISA + Voy} can detect them to 1.4 Gpc. These networks can also constrain black holes in the mass-gap $(4~M_\odot)$ orbiting around a lower-range IMBH $(150~M_\odot)$ within 100 Mpc. But a Pop-B source with upper-range IMBHs, e.g,  ($10^4+220)~M_\odot$, can be detected only by \texttt{LISA + ET} out to $230$ Mpc. The tight bounds on mass-ratios with multiband detection of Pop-B sources would be provide critical constrains on the growth of \imbh{s} in dense stellar environments (e.g., galactic nuclei) through subsequent mergers\cite{2004Natur.428..724P, 2019BAAS...51c.175B}.

\paragraph{ Pop-C (LIGO-like binaries).} In these binaries, the primary object could be a remnant $(\lesssim 100~M_\odot$) of an earlier stellar-origin binary black hole merger, whose one more round of merger with a stellar black hole would lead to lower-range \imbh{s}. In context of multiband observations, these sources are similar to GW150914, i.e. their merger $S/N$ in ground-based detectors will be much larger than the accumulated $S/N$ from the inspiral observed by LISA (starting $f\gtrsim10$ milli-Hz). Thus, their multiband detection radius is limited by the sensitivity of LISA and not by the choice of ground-based detectors. E.g. a second-generation merger of a GW170729-like remnant ($100~M_\odot$) with a $20~M_\odot$ stellar-mass black hole would only have a multiband detection radius of 400 Mpc. While Pop-C would have a lower detection rate (compared to Pop-A,B), constraints from their multiband observations, especially on individual black hole spins, would shed light on formation of compact objects in the pair-instability gap\cite{2019arXiv190605295G}. Instead of early alerts from LISA, the multiband detection of such sources will often be retroactive, i.e., the source parameters supplied by the ground-based detectors will allow researchers go back into archived LISA data and confidently detect Pop-C inspirals that had originally been sub-threshold\cite{2018arXiv180808247W}.

For all mass-ranges of \imbh{s} in Pop-A,B,C classification, clearly \texttt{LISA + ET} is the most sensitive of the networks we consider. However, even in the pessimistic scenario of no 3G detectors being operational by the time LISA flies, the network of \texttt{LISA + Voy} can still lead to strong multiband observations for \imbh{} binaries within total-mass $\lesssim 2000~M_\odot$. By the same token, a non-detection of \imbh{s} by  \texttt{LISA + Voy} would significantly constrain their merger rates.  It would be useful for future studies to extend our analyses to precessing spin configurations and binaries with significant orbital eccentricity.

 \section*{Acknowledgments}
 We would like thank Laura Cadonati and Christopher Berry for useful discussions. We are thankful to Lionel London and Sebastian Khan for the links to generate their waveform model in the LIGO Analysis Library. K.J.'s and D.S.'s research was funded by the NASA Grant 80NSSC19K0322 and the NSF Grant PHY 1806580 and 1809572. C.C.’s work was carried out at the Jet Propulsion Laboratory, California Institute of Technology, under contract to the National Aeronautics and Space Administration. C.C. also gratefully acknowledges support from NSF Grant PHY-1708212. Copyright 2019.  All rights reserved. All the authors have contributed equally to the text and the primary results.


\newpage 
\bibliographystyle{naturemag}

\bibliography{references}

\newpage 
\section*{Supplemental Material}

\paragraph{IMBHs and ground-based detectors.} During epoch-1 (2020s), the mergers of lower-range \imbh{s} will be surveyed out to redshift $z \ sim 2$. The epoch-1 ground network consists of detectors in the U.S. (LIGO) network, India (LIGO), Italy (Virgo) and Japan (KAGRA)\cite{ObservingScenarios}. At their design sensitivity these detectors will be sensitive to gravitational waves down to $10~\text{Hz}$. This implies that the signals from \imbh{} binaries in epoch-1 are very short-lived ($\sim10$ milliseconds), making it difficult to recover their intrinsic properties such as masses and spins\cite{2015PhRvL.115n1101V}. 

In epoch-2 (mid-2020s), we would survey mergers of medium-range \imbh{s} to $z\sim 1$ and lower-range masses to $z\sim 8$. This epoch should see two phases of upgrades to the LIGO facility\cite{LIGO-Instrument2018} - A+, which we find increases the detection redshift of \imbh{s} by 50\% compared to epoch-1, and LIGO-Voyager (\texttt{Voy}), which will push the low-frequency sensitivity envelope down to $5~\text{Hz}$. \texttt{Voy} will extend the sensitive mass-range of \imbh{} binaries up to $\sim 5000~M_\odot$. By epoch-3 (tentatively early to mid-2030), we will survey lower-range (medium-range) \imbh{} mergers out to $z\sim 30$ ($z\lesssim7$). This epoch may consist of two third-generation detectors - Einstein Telescope (\texttt{ET}, a triangular underground detector with 10 km arms)\cite{Punturo:2010zz} and Cosmic Explorer (\texttt{CE}, 40 km arrays)\cite{Evans:2016mbw}. By extending the low-frequency sensitivity range to  $1~\text{Hz}$, \texttt{ET} would provide the first observational glimpse into the population of upper-range \imbh{} binaries. While binaries with lower to medium-range \imbh{s} have strong detection prospects with the epoch-2,3 detectors, recent studies\cite{2016arXiv161006917V} have noted that the bounds on their individual spins remain fairly poor due to degeneracies. This is primarily because the duration of inspiral recorded in the ground-based detectors for \imbh{} sources is not long enough to record a full cycle of spin-orbit precession. This makes it difficult to distinguish between different possible formation channels.

\paragraph{IMBHs and space-based detectors. }
The detection for the upper-range of \imbh{} binaries becomes possible only in epoch-4 (by 2034), when space-based missions such as LISA launch. The detection redshift of binary black holes for a LISA-like satellite constellation depends on the mission lifetime. In this paper, we consider a 6-link LISA (\texttt{L3}) with a nominal 4-year mission lifetime, and we account for noise from the galactic white-dwarf background in addition to instrumental noise (see Methodology for details). We find that LISA would detect the three coalescence stages for upper-range \imbh{} binaries - inspiral, merger, and ringdown - out to $z\sim 500$. It is interesting to note that no other type of discrete astronomical object, in any form of observation, has had the potential to be surveyed so early in the universe.

\begin{figure}
\centering
 \includegraphics[scale=0.34,trim = {50 0 0 0}]{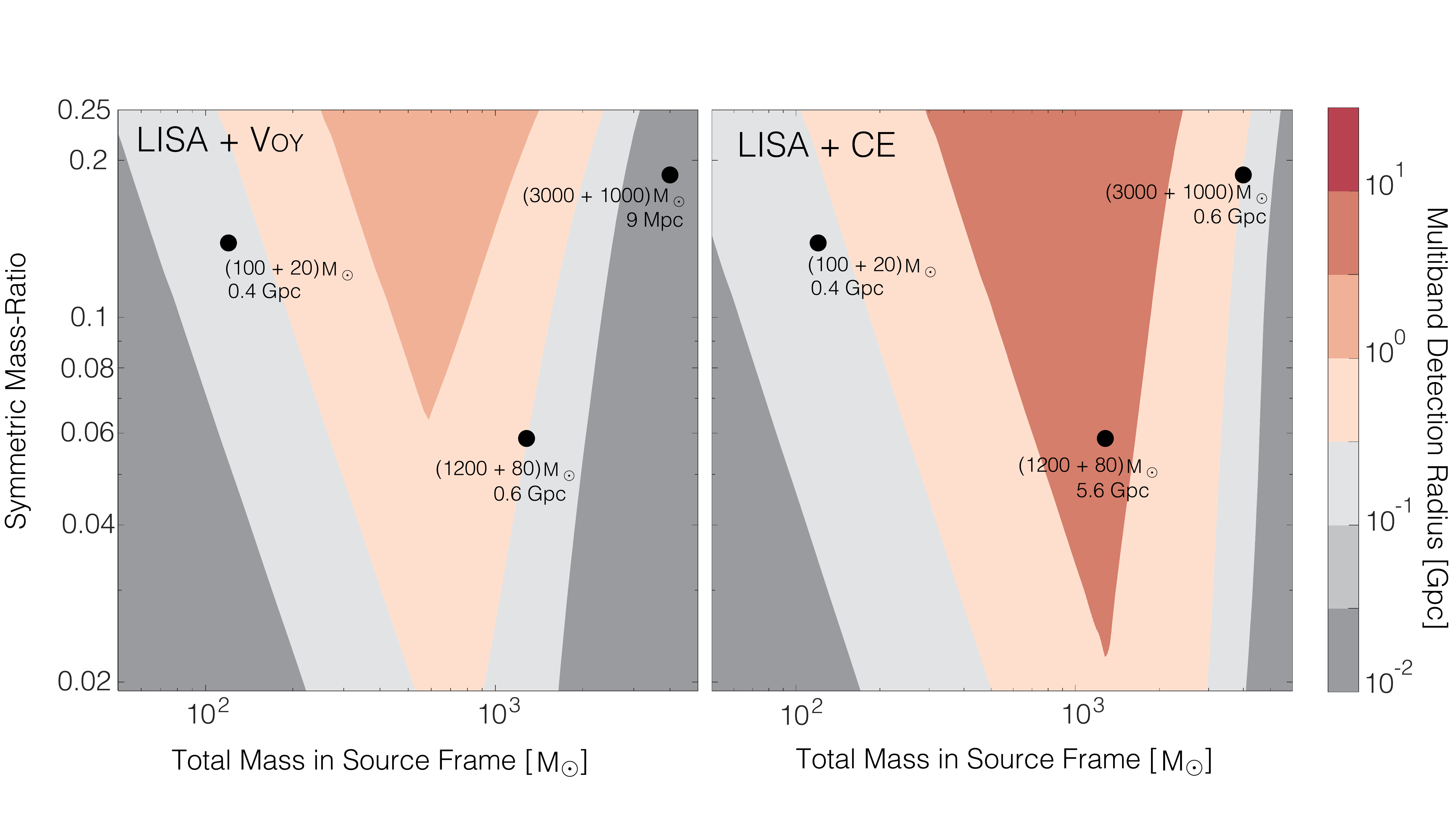} 
\caption{{\bf Multiband parameter space with LISA, \texttt{Voy} and \texttt{CE}}. Similar to figure-\ref{Fig4}, the contours refer to the angle-averaged multiband distance for the \texttt{LISA+Voy} network (right) and  \texttt{LISA+CE} network (left).  }
\label{Fig5} 
\end{figure}

\paragraph{Multiband parameter space for \texttt{LISA + Voy} and \texttt{LISA + CE}. } A binary in either of these network combinations will have a frequency gap of $\Delta{f} \sim 5$ Hz between leaving the LISA band and entering the ground-based band. This low-frequency sensitivity gap for ground-based detectors is principally what limits their multiband detection radius (see figure-\ref{Fig5}). Further, the maximum total mass accessible is $5400~M_\odot$ ($3300~M_\odot$) for \texttt{LISA+CE} (\texttt{LISA + Voy}) network, factor of a few smaller than for \texttt{LISA+ET} ($2\times10^4 M_\odot$).

\begin{figure}
\centering
 \includegraphics[scale=0.4,trim = {80 0 0 0}]{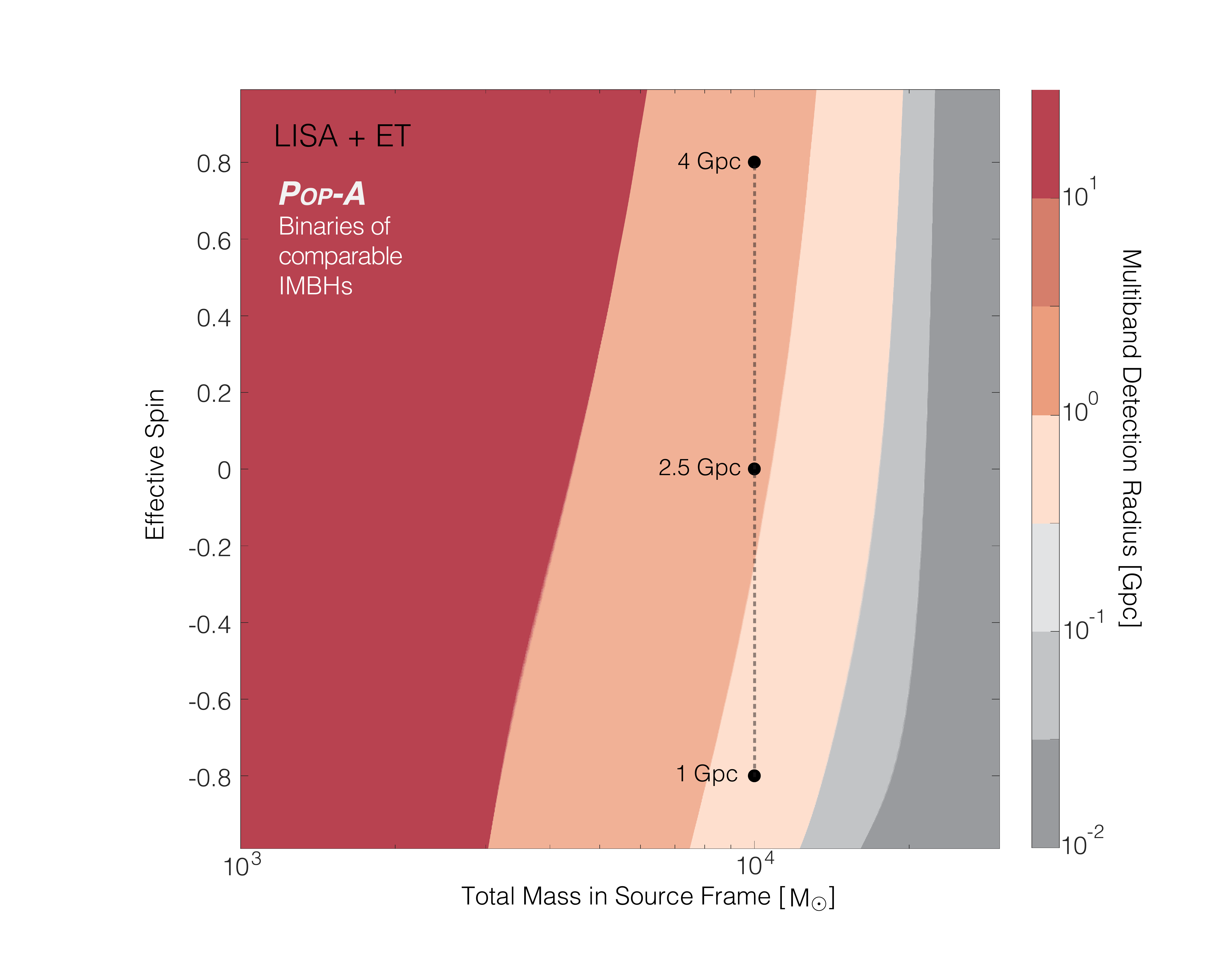} 
\caption{{\bf Impact of black hole spins on detection radius for the \texttt{LISA + ET} network.} The y-axis refers to the effective spin for an equal-mass binary black hole system (symmetric mass-ratio $\eta= 0.25$). The x-axis is the total-mass in the source frame and the contours are angle-averaged multiband distances. The dotted line shows the impact of spins on the  detection radius of $(5000+5000)~M_\odot$ binary.  } 
\label{Fig6} 
\end{figure}

\paragraph{Impact of black hole spins on multiband detection radius.} For the \texttt{LISA + ET} network, we computed the multiband detection radius for varying black hole spins (see figure-\ref{Fig6}). The spins of both the black holes $(\vec{\chi}_{1,2})$ have been combined into a single effective-spin parameter: $\chi_\text{eff} = (\chi_1/m_1 + \chi_2/m_2)\cdot \vec{L} \in [-1,1]$.  Spin precession effects are not included here.
The spins have maximum impact for equal-mass black holes. Therefore, we consider only $m_1/m_2=1$ in figure-\ref{Fig6} and compute the multiband radius at multiple values of total-mass and  $\chi_\text{eff}$. We find that including $\chi_\text{eff}$ has rather limited impact on detectability for 
stellar-mass and lower-range \imbh{} binaries. But for total-mass $\gtrsim 10000~M_\odot$, the spins can change the detection radius by up to $\sim75\% $ (from 1 Gpc to 4 Gpc for alignment of black holes spins with respect to orbital angular momentum).

\newpage
\section*{Methods}

\paragraph{Distances in gravitational wave astronomy. } The luminosity distance $D_L$ to which we can confidently detect a black hole binary with gravitational wave detectors can be expressed as follows:
\begin{eqnarray}
D_L~(S/N) = \mathcal{F} (\Lambda,\ \Theta,\ S_n)
\label{eq-D}
\end{eqnarray}
where $\Lambda$ refers to the intrinsic parameters of the two black holes: masses $(m_1, m_2)$ and spins $(\vec{\chi}_1, \vec{\chi}_2)$. The four extrinsic angles on which the waveform depends (sky-location, orbital orientation) are referred to collectively as $\Theta$, and $S_n$ is the single-sided noise spectral density of the gravitational wave detector. 
Here, $\mathcal{F}$ is implicitly defined by eq. \ref{eq-rho}. It determines the luminosity distance $D_L$ for a gravitational waveform $h(\Lambda, \Theta, D)$ at a fixed value of $S/N$: 
\begin{eqnarray}
S/N = 2~\sqrt{ \int_{f_\text{min}}^{f_\text{max}}~\frac{|~h^2(f; \Lambda, \Theta, D_L)~|}{S_n(f)}~ \text{d}f}
\label{eq-rho}
\end{eqnarray}

The choice of $f_\text{min}$ is generally determined by the noise spectrum of the detector. Since the signal measured by the detector is redshifted by the expansion of universe, the black hole masses $(m_1, m_2)$ in $\Lambda$ have to be scaled with $(1+z(D_L))$. Here $z$ is the cosmological redshift and in this paper $z(D_L)$ is computed for a flat $\Lambda$CDM cosmology, using Planck 2018 results\cite{2018arXiv180706209P}.

The {\bf horizon distance} $z_\text{hor}$ is shown as the y-axis of figure-\ref{Fig2}). It is obtained from eq. \ref{eq-D} for $\Theta$ set to an optimal sky-location and inclination. For ground-based detectors, this refers to a source directly above the detector. 
For computing the horizon distance, the $S/N$ from eq. \ref{eq-rho} is set to 8, which is our nominal value for the detection threshold.

The {\bf multiband detection radius} $D_\text{multi}$ (as luminosity distance) or $z_\text{multi}$ (as redshift) is shown as the y-axis of figure-\ref{Fig3}, contours in figure-\ref{Fig4} and left-hand-side of eq. \ref{eq-multiband}. We obtain this by averaging eq. \ref{eq-D} for all combination of $\Theta$ (sky-location and orientation) and requiring $S/N\geq 8$ in {\it both ground- and space-based detectors}. Our reason to plot this conservative angle-averaged distance, compared to the horizon distance, is that it permits comparison of LISA and ground-based sensitivities on more equal terms. 
The angle averaging brings in an overall factor $\mathcal{C}$ to the $S/N$ value given in eq. \ref{eq-rho}. For all ground-based detectors $\mathcal{C} = 2/5$, while for LISA we use $\mathcal{C} = 2/\sqrt{5}$\cite{Cornish:2018dyw}. (The averaging factor for LISA is different than for ground-based detectors because the LISA community generally follows a different convention when defining $S_n(f)$; here we use the LISA noise formula from\cite{Cornish:2018dyw}, which {\it already} includes (i) factors that incorporate the instrument response averaged over detector orientations, and (ii) the fact that, for our purposes, 3-armed LISA can effectively be treated as two 2-arm IFOs.)
Therefore, the multiband detection radius only depends on the binary's total-mass in the source frame ($M_\text{src}$), the  mass-ratio ($m_1/m_2>1)$, effective spins ($\chi_\text{eff}$) and the minimum $S/N$ measured in {\it both} space- and ground-based detectors. If one chooses horizon distance as a measure for multiband detection radius, it could push the high-mass limit for ground-based detectors. For $1.5\times 10^4 ~M_\odot$ in \texttt{ET}, we find an increase of $\sim 30\%$ (from 0.25 Gpc to 0.38 Gpc) from angle-averaged to optimal distance.

\paragraph{Gravitational Waveform. } We have used a combination of numerical relativity waveforms from the Georgia Tech Catalog \cite{JANI_2016} and a phenomenological waveform model \texttt{IMRPhenomHM} \cite{PhysRevLett.120.161102}. The Georgia Tech simulations are publicly available at \url{einstein.gatech.edu/catalog/}, and the IMRPhenomHM waveforms were generated using the LIGO Analysis Library Suite (\url{https://git.ligo.org/lscsoft/lalsuite}). The numerical relativity waveform includes the sub-dominant harmonics, which are: 
\begin{eqnarray}
(\ell_{NR}, m_{NR}) = \left\lbrace(2,\pm2), (2,\pm1), (3,\pm2), (3,\pm3), (4,\pm3), (4,\pm4), (5,\pm5), (6,\pm6) \right\rbrace 
\end{eqnarray}
However, the IMR model includes fewer modes: 
\begin{eqnarray}
(\ell_{model}, m_{model}) =\left\lbrace(2,\pm2),(2,\pm1),(3,\pm2),(3,\pm3),(4,\pm3),(4,\pm4)\right\rbrace 
\end{eqnarray}
The motivation for combining two different waveform-generating techniques is to capture all essential physics, both in the post-Newtonian and extreme gravity regimes.

\paragraph{Instrumental Noise. } For ground-based detectors, the frequency limits in eq. (2) depend on the choice of observatory. As we are interested stellar-maass and intermediate-mass black hole binaries, the $f_\mathrm{max}$ for all ground-based detectors can be kept at $1024~\mathrm{Hz}$.  However the lower limit $f_mathrm{min}$ for ground-based detectors should decrease substantially from now (Advanced LIGO runs in 2015-2019) until the time when LISA flies (2034), and $f_\mathrm{min}$ significantly impacts the detection radius. For Figure \ref{Fig2} the instrumental noise $S_n$ for the current generation of ground-based observatories - Advanced LIGO (aLIGO), Advanced Virgo (aVirgo) and KAGRA - is set to their design sensitivity $(f_\mathrm{min} =10~\mathrm{Hz})$, which could be achieved by 2022 \cite{ObservingScenarios}. LIGO-Voyager is expected to be sensitive down to $f_\mathrm{min} =5~\mathrm{Hz}$. Meanwhile, new facilities such as Einstein Telescope and Cosmic Explorer should observe gravitational waves with frequencies down to $f_\mathrm{min} =1~\mathrm{Hz}$ and $5~\mathrm{Hz}$ respectively. 

For space-based missions like LISA, the frequency range goes from $f_\mathrm{min}=0.1~\mathrm{mHz}$ and $f_\mathrm{max}=1~\mathrm{Hz}$. For our study, the instrumental noise of LISA corresponds to the L3 design (three satellites separated by 2.5 million kms), details regarding which are stated in a recent review\cite{Cornish:2018dyw}. As the 6-link LISA-L3 configuration is essentially a network of two detectors, the noise curve we have utilized already includes a factor of $\sqrt{2}$. It also includes the effect averaging over sky-location and LISA's solar orbit. While we have also included the noise from the galactic white dwarf binaries in LISA (causes a drop in sensitivity after the peak in the figure-\ref{Fig2}), it has no impact on multiband observations. As LISA's mission lifetime is only 4-years, any binary of intermediate and stellar masses will spend only a tiny fraction in the frequency range. Therefore, while computing $S/N$, we have ensured the frequency range corresponds to coalescence time of less than 4 years. Also, in computing the horizon distances in figure-\ref{Fig1}, we multiplied eq.-\ref{eq-rho} by $\sqrt{5}$ to account for optimal orientation.  

The noise curves for the ground-based detectors used in this study are available at \url{https://dcc.ligo.org/LIGO-T1500293/public}. The noise curve for the LISA detector is computed using procedure described in recent text \cite{Cornish:2018dyw}.  

\paragraph{Derivation of Forumula for Multiband Detection Radius. } In eq. \ref{eq-multiband}, we provide an analytic expression for the multiband detection radius $z_\text{multi}$ for an equal-mass, non-spinning multiband binary source, for each of the three networks. This expression depends only on the total mass measured in the detector frame, $M_\text{det} = M_\text{src} (1 + z_{\text{multi}} )$, and the threshold value of $S/N$ (for both ground- and space-based detectors). 

Our expression is divided into three separate mass bins: (a) Stellar-mass binaries, in which $M_\text{det}\in[10, 100)$, (b) lower-range IMBH binaries, in which $M_\text{det}\in[10, M_0)$ and (c) medium-range IMBH binaries, where $M_\text{det}\geq M_0$. The value of $M_0$ corresponds to the mass of the binary that can be seen the farthest in multiband observations, given the detection threshold of $S/N=8$.

The constants $a_i, b_i$ depend on the noise spectrum of the space-detector. For the L3 configuration of LISA, 
\begin{eqnarray}
a_1 = 0.09063,~a_2=1.65, \\
b_1 = 1.478,~b_2=0.6576,~b_3=-0.2769 
\label{constants}
\end{eqnarray}
The constants $c_i$ and $ M_0$, listed in table-\ref{Table-p}, depend on the noise spectrum of the ground-based detectors. For binaries with stellar-mass black holes and lower-mass range \imbh{s}, the multiband detection radius is limited by the sensitivity of LISA. For medium-range \imbh{s}, the radius is limited mostly by the low-frequency limits of the ground-based detectors.

\begin{table}
\centering
\begin{tabular}{ |p{4cm}||p{1.5cm}|p{1.5cm}|p{1.5cm}|p{1.5cm}|p{1.5cm}|  }
 \hline
Detector    & $M_0$ & $c_1$ & $c_2$ \\
 \hline
 Einstein Telescope  & 6650 &   59.84 & -2.451\\
 Cosmic Explorer   & 3151 &  667.9 &  -5.462 \\
LIGO-Voyager & 1393&   21.7 &   -2.534 \\
\hline
\end{tabular}
\caption{Constants for eq. \ref{eq-multiband}c }
\label{Table-p}
\end{table}


\end{document}